\documentclass{aastex631}
\usepackage{graphicx}
\usepackage[caption=false, labelformat=empty]{subfig}
\usepackage{comment}

\shorttitle{Simons Observatory focal-plane module design}
\shortauthors{McCarrick et al.}

\graphicspath{{./}{figs/}}

\begin{document}

\title{The Simons Observatory microwave SQUID multiplexing detector module design}

\email{hm8@princeton.edu}

\author{Heather McCarrick}
\affiliation{Joseph Henry Laboratories of Physics, Jadwin Hall, Princeton University, Princeton, NJ 08544, USA } 

\author{Erin Healy}
\affiliation{Joseph Henry Laboratories of Physics, Jadwin Hall, Princeton University, Princeton, NJ 08544, USA }

\author[0000-0002-9957-448X]{Zeeshan Ahmed}
\affiliation{Kavli Institute for Particle Astrophysics and Cosmology, Menlo Park, CA 94025, USA}  
\affiliation{SLAC National Accelerator Laboratory, Menlo Park, CA 94025, USA}

\author{Kam Arnold}
\affiliation{Department of Physics, University of California San Diego, La Jolla, CA 92093, USA}

\author[0000-0002-2287-1603]{Zachary Atkins}
\affiliation{Joseph Henry Laboratories of Physics, Jadwin Hall, Princeton University, Princeton, NJ 08544, USA }

\author{Jason E. Austermann}
\affiliation{National Institute of Standards and Technology, Boulder, CO 80305, USA}

\author[0000-0002-2971-1776]{Tanay Bhandarkar}
\affiliation{Department of Physics and Astronomy, University of Pennsylvania, Philadelphia, PA 19104, USA}

\author{James A. Beall}
\affiliation{National Institute of Standards and Technology, Boulder, CO 80305, USA}

\author{Sarah Marie Bruno}
\affiliation{Joseph Henry Laboratories of Physics, Jadwin Hall, Princeton University, Princeton, NJ 08544, USA }

\author[0000-0002-9113-7058]{Steve K. Choi}
\affiliation{Department of Physics, Cornell University, Ithaca, NY 14853, USA}
\affiliation{Department of Astronomy, Cornell University, Ithaca, NY 14853, USA}

\author{Jake Connors}
\affiliation{National Institute of Standards and Technology, Boulder, CO 80305, USA}

\author{Nicholas F. Cothard}
\affiliation{Department of Applied and Engineering Physics, Cornell University, Ithaca, NY 14853, USA}

\author{Kevin D. Crowley}
\affiliation{Joseph Henry Laboratories of Physics, Jadwin Hall, Princeton University, Princeton, NJ 08544, USA }

\author[0000-0002-1940-4289]{Simon Dicker}
\affiliation{Department of Physics and Astronomy, University of Pennsylvania, Philadelphia, PA 19104, USA}

\author[0000-0002-6817-0829]{Bradley Dober}
\affiliation{Department of Physics, University of Colorado, Boulder, CO 80309, USA}
\affiliation{National Institute of Standards and Technology, Boulder, CO 80305, USA}

\author{Cody J. Duell}
\affiliation{Department of Physics, Cornell University, Ithaca, NY 14853, USA}

\author{Shannon M. Duff}
\affiliation{National Institute of Standards and Technology, Boulder, CO 80305, USA}

\author[0000-0002-9962-2058]{Daniel Dutcher}
\affiliation{Joseph Henry Laboratories of Physics, Jadwin Hall, Princeton University, Princeton, NJ 08544, USA }

\author{Josef C. Frisch}
\affiliation{SLAC National Accelerator Laboratory, Menlo Park, CA 94025, USA}

\author[0000-0001-7225-6679]{Nicholas Galitzki}
\affiliation{Department of Physics, University of California San Diego, La Jolla, CA 92093, USA}

\author{Megan B. Gralla}
\affiliation{Department of Astronomy/Steward Observatory, University of Arizona, 933 N. Cherry Ave., Tucson, AZ 85721, USA}

\author[0000-0003-1760-0355]{Jon E. Gudmundsson}
\affiliation{The Oskar Klein Centre, Department of Physics, Stockholm University, SE-106 91 Stockholm, Sweden}

\author[0000-0001-7878-4229]{Shawn W. Henderson}
\affiliation{Kavli Institute for Particle Astrophysics and Cosmology, Menlo Park, CA 94025, USA}  
\affiliation{SLAC National Accelerator Laboratory, Menlo Park, CA 94025, USA}

\author{Gene C. Hilton}
\affiliation{National Institute of Standards and Technology, Boulder, CO 80305, USA}

\author{Shuay-Pwu Patty Ho}
\affiliation{Department of Physics, Stanford University, Stanford, CA 94305 USA}

\author[0000-0003-4573-4094]{Zachary B. Huber}
\affiliation{Department of Physics, Cornell University, Ithaca, NY 14853, USA}

\author{Johannes Hubmayr}
\affiliation{National Institute of Standards and Technology, Boulder, CO 80305, USA}

\author[0000-0001-7466-0317]{Jeffrey Iuliano}
\affiliation{Department of Physics and Astronomy, University of Pennsylvania, Philadelphia, PA 19104, USA}

\author[0000-0002-6898-8938]{Bradley R. Johnson}
\affiliation{Department of Astronomy, University of Virginia, Charlottesville, VA 22904, USA}

\author[0000-0001-5374-1767]{Anna M. Kofman}
\affiliation{Department of Physics and Astronomy, University of Pennsylvania, Philadelphia, PA 19104, USA}

\author{Akito Kusaka}
\affiliation{Physics Division, Lawrence Berkeley National Laboratory, Berkeley, CA 94720, USA}
\affiliation{Department of Physics, The University of Tokyo, Tokyo 113-0033, Japan}
\affiliation{Kavli Institute for the Physics and Mathematics of the Universe (WPI), Berkeley Satellite, University of California, Berkeley 94720, USA}
\affiliation{Research Center for the Early Universe, School of Science, The University of Tokyo, Tokyo 113-0033, Japan}

\author[0000-0002-6522-6284]{Jack Lashner}
\affiliation{Department of Physics and Astronomy, University of Southern California, Los Angeles, CA 90089, USA}

\author[0000-0003-3106-3218]{Adrian T. Lee}
\affiliation{Department of Physics, University of California Berkeley, Berkeley, CA 94720, USA}
\affiliation{Physics Division, Lawrence Berkeley National Laboratory, Berkeley, CA 94720, USA}
\affiliation{Radio Astronomy Lab, University of California Berkeley, Berkeley, CA 94720, USA}

\author[0000-0001-8093-2534]{Yaqiong Li}
\affiliation{Department of Physics, Cornell University, Ithaca, NY 14853, USA}
\affiliation{Kavli Institute at Cornell for Nanoscale Science, Cornell University, Ithaca, NY 14853, USA}

\author{Michael J. Link}
\affiliation{National Institute of Standards and Technology, Boulder, CO 80305, USA}

\author{Tammy J. Lucas}
\affiliation{National Institute of Standards and Technology, Boulder, CO 80305, USA}

\author{Marius Lungu}
\affiliation{Joseph Henry Laboratories of Physics, Jadwin Hall, Princeton University, Princeton, NJ 08544, USA }

\author{J.A.B. Mates}
\affiliation{National Institute of Standards and Technology, Boulder, CO 80305, USA}

\author{Jeffrey J.  McMahon}
\affiliation{Department of Astronomy and Astrophysics, University of Chicago, 5640 S. Ellis Ave., Chicago, IL 60637 USA}
\affiliation{USA Kavli Institute for Cosmological Physics, University of Chicago, 5640 S. Ellis Ave., Chicago, IL 60637, USA}

\author[0000-0001-7125-3580]{Michael D. Niemack}
\affiliation{Department of Physics, Cornell University, Ithaca, NY 14853, USA}
\affiliation{Department of Astronomy, Cornell University, Ithaca, NY 14853, USA}
\affiliation{Kavli Institute at Cornell for Nanoscale Science, Cornell University, Ithaca, NY 14853, USA}

\author[0000-0003-1842-8104]{John Orlowski-Scherer}
\affiliation{Department of Physics, University of California San Diego, La Jolla, CA 92093, USA}

\author[0000-0002-0298-9911]{Joseph Seibert}
\affiliation{Department of Physics, University of California San Diego, La Jolla, CA 92093, USA}

\author[0000-0001-7480-4341]{Maximiliano Silva-Feaver}
\affiliation{Department of Physics, University of California San Diego, La Jolla, CA 92093, USA}

\author{Sara M. Simon}
\affiliation{Fermi National Accelerator Laboratory, Batavia, IL 60510, USA}

\author{Suzanne Staggs}
\affiliation{Joseph Henry Laboratories of Physics, Jadwin Hall, Princeton University, Princeton, NJ 08544, USA }

\author[0000-0001-8101-468X]{Aritoki Suzuki}
\affiliation{Physics Division, Lawrence Berkeley National Laboratory, Berkeley, CA 94720, USA}

\author[0000-0002-2380-0436]{Tomoki Terasaki}
\affiliation{Department of Physics, The University of Tokyo, Tokyo 113-0033, Japan}

\author{Joel N. Ullom}
\affiliation{National Institute of Standards and Technology, Boulder, CO 80305, USA}

\author{Eve M. Vavagiakis}
\affiliation{Department of Physics, Cornell University, Ithaca, NY 14853, USA}

\author{Leila R. Vale}
\affiliation{National Institute of Standards and Technology, Boulder, CO 80305, USA}

\author{Jeff Van Lanen}
\affiliation{National Institute of Standards and Technology, Boulder, CO 80305, USA}

\author{Michael R. Vissers}
\affiliation{National Institute of Standards and Technology, Boulder, CO 80305, USA}

\author{Yuhan Wang}
\affiliation{Joseph Henry Laboratories of Physics, Jadwin Hall, Princeton University, Princeton, NJ 08544, USA }

\author[0000-0002-7567-4451]{Edward J. Wollack}
\affiliation{NASA Goddard Space Flight Center, Greenbelt, MD 20771, USA}

\author[0000-0001-5112-2567]{Zhilei Xu}
\affiliation{Department of Physics and Astronomy, University of Pennsylvania, Philadelphia, PA 19104, USA}
\affiliation{MIT Kavli Institute, Massachusetts Institute of Technology, Cambridge, MA 02139, USA}

\author{Edward Young}
\affiliation{Kavli Institute for Particle Astrophysics and Cosmology, Menlo Park, CA 94025, USA}  
\affiliation{Department of Physics, Stanford University, Stanford, CA 94305, USA}

\author{Cyndia Yu}
\affiliation{Kavli Institute for Particle Astrophysics and Cosmology, Menlo Park, CA 94025, USA}  
\affiliation{Department of Physics, Stanford University, Stanford, CA 94305 USA}

\author{Kaiwen Zheng}
\affiliation{Joseph Henry Laboratories of Physics, Jadwin Hall, Princeton University, Princeton, NJ 08544, USA }

\author{Ningfeng Zhu}
\affiliation{Department of Physics and Astronomy, University of Pennsylvania, Philadelphia, PA 19104, USA}

\begin{abstract}
Advances in cosmic microwave background (CMB) science depend on increasing the number of sensitive detectors observing the sky. New instruments deploy large arrays of superconducting transition-edge sensor (TES) bolometers tiled densely into ever larger focal planes.  High multiplexing factors reduce the thermal loading on the cryogenic receivers and simplify their design. 
We present the design of focal-plane modules with an order of magnitude higher multiplexing factor than has previously been achieved with TES bolometers. We focus on the novel cold readout component, which employs microwave SQUID multiplexing ($\mu \mathrm{mux}$). Simons Observatory will use 49 modules containing 60,000 bolometers to make exquisitely sensitive measurements of the CMB. 
We validate the focal-plane module design, presenting measurements of the readout component with and without a prototype detector array of 1728 polarization-sensitive bolometers coupled to feedhorns. The readout component achieves a 95\% yield and a 910 multiplexing factor. The median white noise of each readout channel is 65~$\mathrm{pA/\sqrt{Hz}}$. This impacts the projected SO mapping speed by $< 8\%$, which is less than is assumed in the sensitivity projections. The results validate the full functionality of the module. We discuss the measured performance in the context of SO science requirements, which are exceeded. 

\end{abstract}

\section{Introduction} \label{sec:intro}

The Simons Observatory (SO) is a suite of new telescopes that will be used to survey the cosmic microwave background (CMB) temperature and polarization anisotropies.
These observations are designed to measure or constrain the primordial B-mode polarization signal with a tensor-to-scalar ratio $r$ uncertainty of $\sigma(r) = 0.003$, constrain the sum of the neutrino masses, and survey galaxy clusters through the thermal and kinematic Sunyaev-Zel'dovich effects, along with a number of other cosmological and astrophysical probes~\citep{so_forecast_2019}. 

The initial configuration of SO consists of one large 6~m aperture telescope (LAT)~\citep{niemack_2015, Gudmundsson2021, xu2021} and three small aperture 0.5~m telescopes (SATs)~\citep{aamir_2019,galitzki_2018}.
The four telescopes will observe in six common spectral bands centered between 40 and 290~GHz. 
Collectively, dichroic pixels with over 60,000 transition-edge-sensor (TES) bolometers will be deployed to achieve the sensitivity needed for the cosmological science goals. 
The optical coupling, detector array, and cold readout are packaged in a universal focal-plane module (UFM). 
Each SAT focal plane is comprised of seven tiled UFMs. 
The LAT receiver (LATR)~\citep{zhu_2021} will contain seven optics tubes, each of which has a focal plane with three UFMs, for a total of 21 UFMs.
The `universal' name is given to the modules as they are common to all receiver focal planes. 
The detector arrays build on the proven detector performance of the Atacama Cosmology Telescope (ACT)~\citep{henderson_2015} and POLARBEAR~\citep{suzuki_2016} collaborations.

The number of detectors SO requires to meet the stated cosmological goals is a significant increase over current experiments, which in turn requires both readout and focal-plane module development. 
Current CMB experiments use voltage-biased TES bolometers noise-limited by photons. Their proven sensitivity across millimeter-wave frequencies makes them the standard.  Thus, any significant gain in sensitivity necessarily comes from an increase in the detector count. 

The readout architectures must advance to accommodate the growing number of TESs and associated circuitry. Simply scaling existing readout designs will result in high thermal loads,  complex cryogenic wiring designs, and/or inefficient use of the focal-plane area; these are at odds with other experiment demands. 
For SO, we are increasing the multiplexing factor to, colloquially, 1000 (in practice, 910) in half an octave of bandwidth. This allows efficient packing of focal-plane modules and significantly decreases the cold wiring complexity as compared to current readout systems~\citep{dobbs2012,dekorte2003}, as all detectors within a module share only two transmission lines. Furthermore, we show this can be implemented in a compact assembly, which can be tessellated, enabling large, densely packed focal planes. The design is compatible with the requirement that the readout and detectors have sufficiently low noise such that the desired instrument sensitivity can be achieved.  

\begin{figure*}[t]
\centering
\includegraphics[%
    width=0.85\linewidth,
  keepaspectratio]{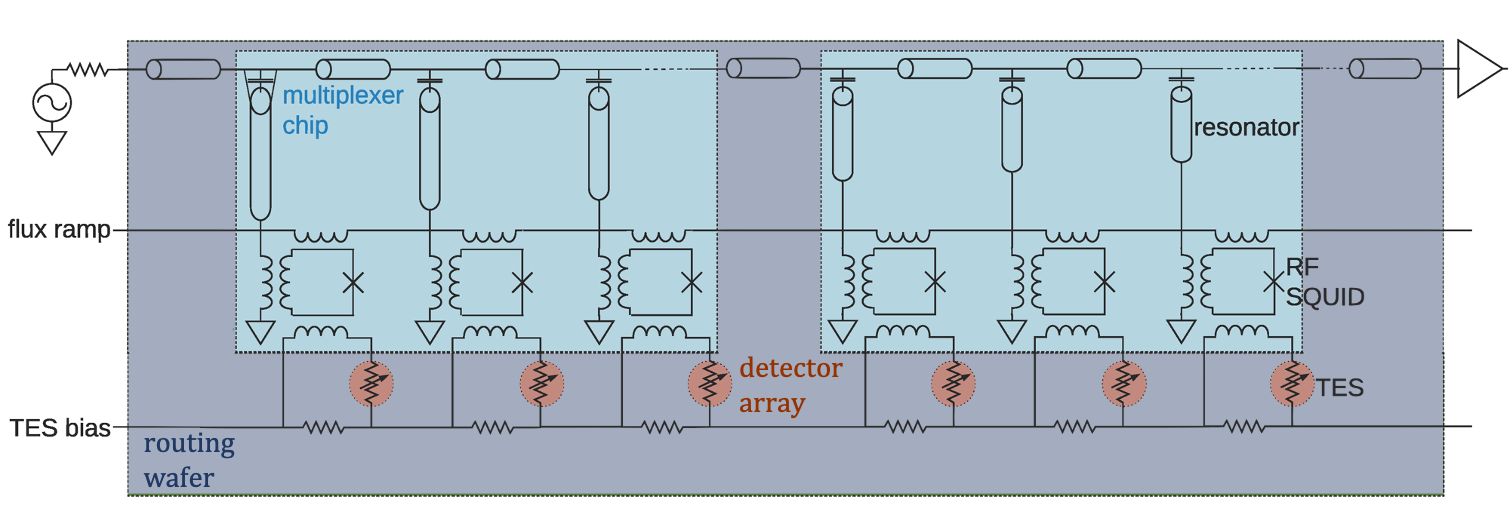}
\caption{Schematic showing the microwave SQUID multiplexing ($\mu \mathrm{mux}$) circuit and implementation in the Simons Observatory focal-plane module. The routing wafer is marked in {\it grey}, multiplexer chips {\it blue}, and TES detector array {\it red}. See~\citet{mates_thesis} for additional details of the on-chip $\mu \mathrm{mux}$ circuit.}
\label{fig:umux_schematic}
\end{figure*}

\subsection{Microwave SQUID multiplexing}

SO will use microwave superconducting quantum interference device (SQUID) multiplexing ($\mu \mathrm{mux}$)~\citep{Irwin04, mates08} to read out the detectors. 
Microwave SQUID multiplexing transforms a change in TES current into a shift in resonance frequency $f_\mathrm{0}$ of a resonator, which is read out through a phase shift of the interrogating probe tone. 
A schematic is shown in Fig.~\ref{fig:umux_schematic}.
Each TES is inductively coupled to a radio frequency (RF) SQUID, which is in turn inductively coupled to a quarter-wave co-planar CPW resonator. 
Each resonator has a different electrical length and thus unique $f_\mathrm{0}$, allowing many channels to be read out on a single transmission line.
The flux ramp is a second input to the SQUID; it continually applies a sawtooth signal, which linearizes the SQUID response and modulates (at $\sim 16$~kHz) the TES signal above any elevated $1/f$ noise, including the two-level system noise frequently seen in microwave resonators~\citep{gao}. 
Microwave multiplexing is a novel readout method for CMB experiments and offers an elegant solution for multiplexing an order of magnitude more detectors than previous architectures. 

In the remainder of this paper, we detail and motivate the design of the readout module in Sec.~\ref{sec:design}, validate the design in Sec.~\ref{sec:validation}, discuss the performance in the context of SO requirements in Sec.~\ref{sec:so_requirements}, and conclude with possible future directions in Sec.~\ref{sec:future}.

\section{Design} \label{sec:design}
\subsection{Readout overview}
The SO readout system consists of three parts: the warm electronics, the readout wiring (RF chains and DC wiring), and the cold (100~mK) multiplexer assembly, the last of which this paper details. For the warm readout (300~K and external to the cryostat), SO uses the  SLAC Superconducting Microresonator Radio Frequency (SMuRF) electronics~\citep{henderson2017}. SMuRF generates the tones which drive, interrogate, and track the resonators and provides the DC TES biases and flux ramp signals (nominally 4~kHz). The probe tones enter and exit the cryostat via the RF chains. The RF chains~\citep{rao2019} contain a series of attenuators on the input side to achieve the desired tone power at the cold multiplexer input. Two-stage amplification (at 4~K and 40~K) on the output, in concert with the tone tracking, achieves high gain with sufficient linearity to maintain low noise at the total power needed for 910 tones. 

\begin{figure*}[t!]
\centering
\includegraphics[%
  width=1\linewidth,
  keepaspectratio]{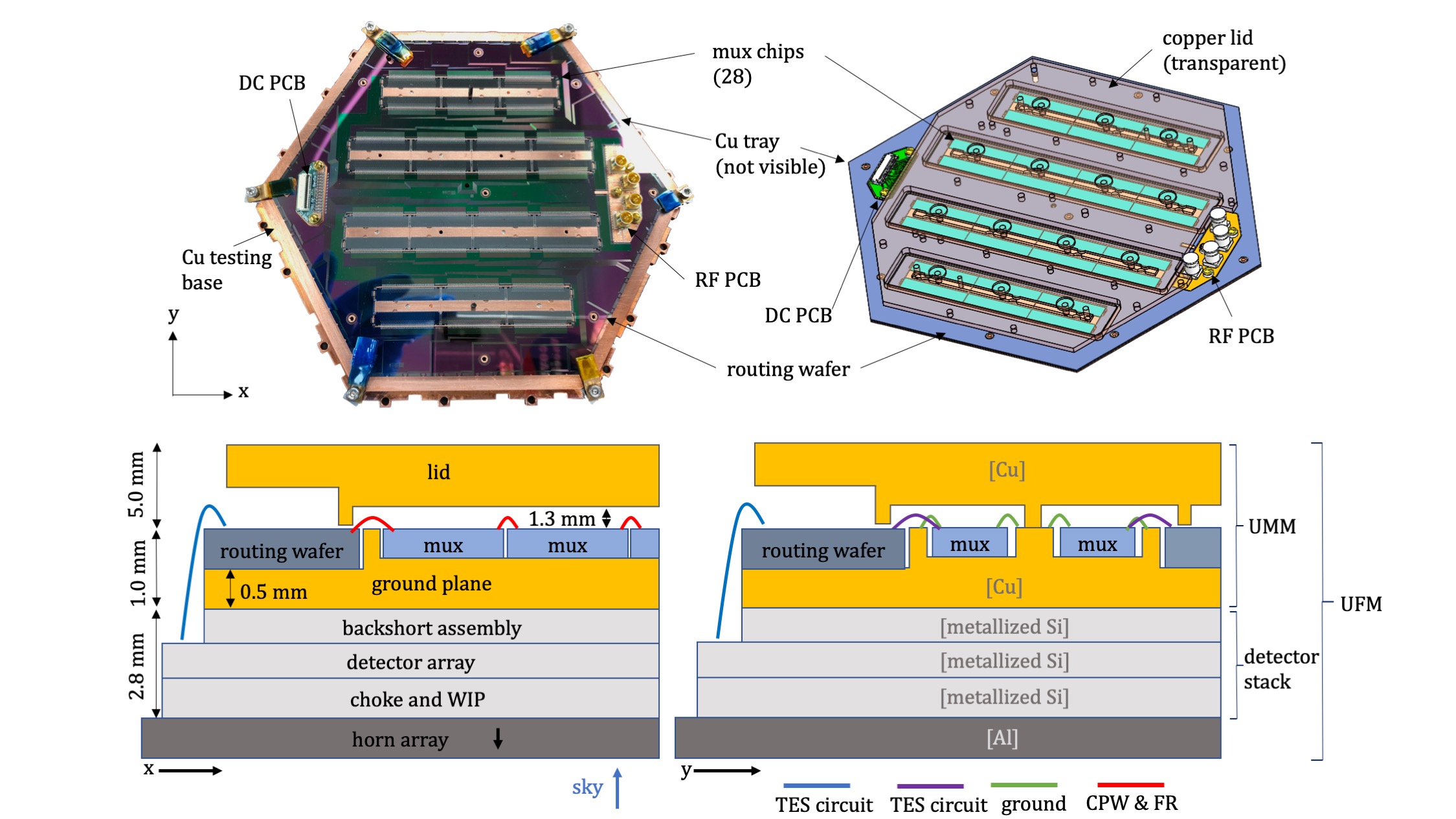}
\caption{
{\it Upper} Photograph and rendering of the readout module, which contains the 100~mK readout components. The lid is removed to reveal the silicon components, including the two sets of 14~multiplexer chips, containing the 1820 readout channels, and the routing wafer, which connects the multiplexer chips in series and hosts the TES bias circuitry. The silicon components rest on a copper plane, which is the RF ground and mechanical support. The $\mu \mathrm{mux}$ components are independently characterized before being coupled to a detector array. The right rendering shows the copper lid (transparent). 
{\it Lower} Cross-sectional views (not-to-scale) of the universal focal-plane module.
The left view illustrates that rows of multiplexer chips are connected in series at the coplanar waveguide (CPW) and flux ramp with wirebonds, as shown in {\it red}. The right shows a portion of the ground bonds that are made between the multiplexer chips and ground plane as well as the enclosure of the multiplexer chips into cavities. In both views, the electrical connections of the transition-edge sensor (TES) circuit, are achieved through wirebonds. This connection between multiplexer chip and routing water is denoted in {\it purple} and between the routing wafer and detector array in {\it blue}. The approximate z-dimensions are shown in millimeters.
}
\label{fig:module}
\end{figure*}

\subsection{Readout module}

The novel part of the UFM is contained in the readout module, referred to as the universal $\mu \mathrm{mux}$ module (UMM). 
The UMM is a compact, low-profile assembly that sits directly behind the 150~mm diameter detector array, occupying the same footprint.
The UMM is shown in Fig.~\ref{fig:module} and the comprising detector and $\mu \mathrm{mux}$ components are given in Table~\ref{tab:ufm}. 
Two pairs of coaxial lines read out the detectors for a total of 1820 channels in a $2 \times 910$ configuration. 
Twenty-eight multiplexer chips ($2 \times 14$) read out between 4--6~GHz are mounted to a monolithic copper piece that provides both a ground plane and mechanical support. The microwave readout and flux ramp circuits continue on a monolithic silicon routing wafer mounted in the same plane as the multiplexer chips. The routing wafer also provides the 400 $\mu\Omega$ shunt resistors for each TES bias circuit, and the routing for twelve TES bias lines. The shunt resistances were chosen to be chosen to be $10\%$ of the TES operating resistance while the bias line number was informed by the achieved uniformity of the Advanced ACTPol arrays. 
Each detector array contains up to 1728 optically coupled detectors (and 36 dark bolometers). The UFM is operated at a bath temperature of 100~mK.
We discuss the components and design choices in more detail in the remainder of this section, and touch on the assembly process~\citep{healy_2020}, which is an important component in achieving good RF performance.

\begin{figure*}[t]
\centering
\includegraphics[width=0.9\linewidth]{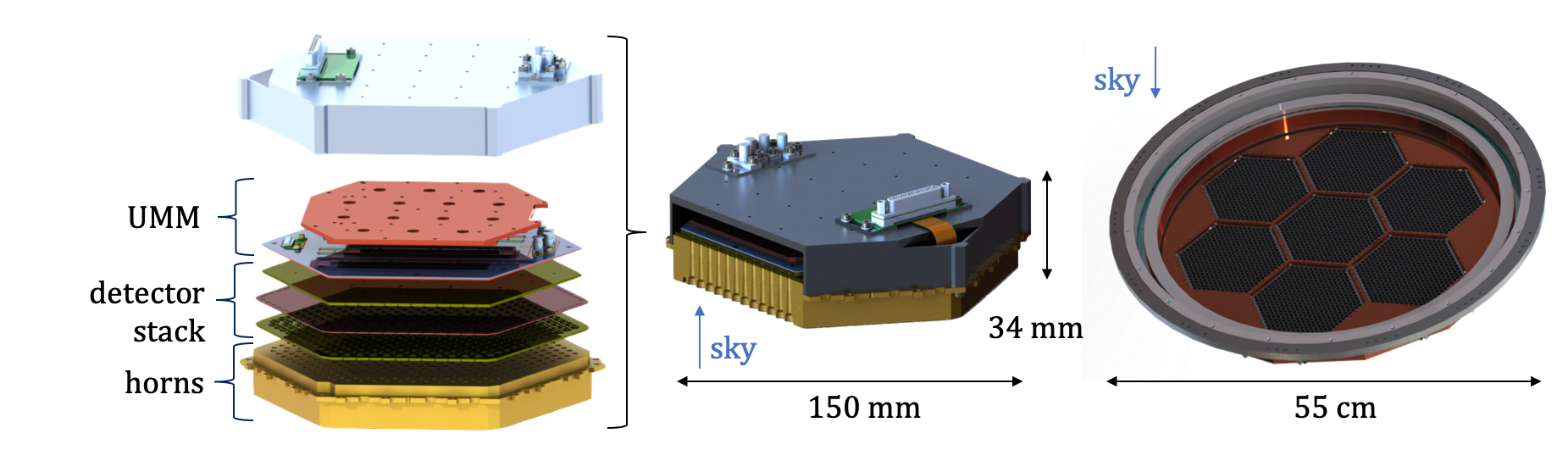}
\caption{
{\it Left} Exploded view of the SO focal-plane module. The UFM is comprised of (i) the UMM {\it blue and copper}, (ii) the detector stack {\it yellow} including the array {\it pink}, and (iii) the horns {\it gold}.  After validation, the UMM is transferred on top of the detector wafer stack. The TES bias line connections between the routing wafer and detector array are made with wirebonds around the perimeters.
{\it Center} Assembled view of the SO UFM. Not pictured is an outer aluminum shield nor the corresponding connectors to the receiver wiring. 
{\it Right} Rendering of the first SAT focal plane with seven detector arrays. Tightly packing the hexagonal detector tiles in the SAT focal planes limits how large the readout module area can be while the LAT focal planes (not shown) limit the height.}
\label{fig3}
\end{figure*}

 A single multiplexer chip contains 66 multiplexer channels, one of which is used for noise diagnostics and one of which terminates at a bare resonator. Briefly, each channel on the chip consists of TES inputs, an RF SQUID, and a quarter-wave resonator with a unique $f_\mathrm{0}$ set by the CPW length. The resonators are capacitively coupled to a common CPW transmission line. The common flux ramp line is inductively coupled to the SQUIDs. The resonators are designed to have a bandwidth $b_\mathrm{r} = 100~\mathrm{kHz}$, which is achieved by tuning the coupling quality factor $Q_\mathrm{c}$ for a given $f_\mathrm{0}$ and internal quality factor $Q_\mathrm{i}$. The coupling here refers to that between the resonator and transmission line. A set of 14 multiplexer chips covers the 4--6~GHz readout bandwidth~\citep{dober_2020}. 

A copper plane physically hosts the multiplexer chips and provides a ground reference for the UMM. This copper plane is 500~$\mu\mathrm{m}$ thick and machined into a hexagon of the same dimensions as the routing wafer. The multiplexer chips are diebonded with rubber cement onto the copper plane using a flip chip bonder.  Extruded features surround rows of either three or four multiplexer chips. These features are paired to those in a copper lid, which mates to the copper plane. This effectively reduces the size of the RF package to reduce the number and coupling to low-frequency box modes, which have been observed to degrade the resonator $Q_\mathrm{i}$ and thus performance. The lid hosts pogo-pins, which push down on and mechanically restrain the routing wafer. 

The copper plane also critically acts as the RF ground and is continuous beneath the multiplexer chips. Early versions used a silicon wafer with a monolithic, deposited ground plane, which compromised the achievable resonator quality factor and motivated the use of a copper ground plane (J.A.B. Mates 2019, private communications). Due to the segmented nature of the assembly -- with 28 multiplexer chips -- it is crucial that all of the ground planes be tied together with a sufficiently low RF impedance. The resonator $Q_\mathrm{i}$ is sensitive to the nearby grounding scheme. The modularity of the multiplexer chips allows for ground connections to the copper plane via wirebonds at each chip very close to each resonator. The grounds of the multiplexer chips and the routing wafer are tied together with wirebonds as well. The connection from the RF ground to the package is made via a PCB; the PCB is connected to the package ground with vias and the routing wafer ground with wirebonds. This PCB transitions the RF signal and ground to the plane of the silicon components via a CPW to surface mount sub miniature push-on (SMP) transition. The RF signals ultimately interface to the receiver wiring through conventional feedthroughs, which mate to the SMP connectors. This architecture effectively brings all necessary electrical lines into the module without using focal plane area.  

The lithographed routing wafer serves the dual purpose of (i) connecting the rows of multiplexer chips in series and (ii) providing the TES bias circuitry. First, the routing wafer has cutouts at the multiplexer chip locations, and thus fits around the rows of multiplexer chips like a frame and rests on the copper plane. The silicon is 500~$\mu$m thick and etched into a hexagonal shape from a 150~mm diameter wafer. Superconducting CPW and flux ramp lines are routed between the multiplexer chip rows on the wafer. Wirebonds connect the flux ramp and CPW lines on the multiplexer chips and routing wafer in series with 18 interconnections for each of the sets of 14 multiplexer chips. The CPW wirebonds are made such that the impedance mismatch ($<5~\Omega$) due to the wirebond inductance is minimized~\citep{li_2019}; a full schedule is shown in~\citet{healy_2020}. The routing wafer also connects the transmission lines of the first and last multiplexer chips to the RF and DC boards that interface with the rest of the cryostat. Second, the routing wafer hosts the TES shunt resistors and bias lines. The TES inputs are  wirebonded from the multiplexer chips to the routing wafer. The bias lines emerge at bond pads at the perimeter of the routing wafer. The TES bias signals interface to the routing wafer via the planar DC PCB board, which is also the launch point for the flux ramp signals. These signals ultimately interface with the receiver wiring through a flex-cable, which ends in a conventional connector (a Micro D-subminiature).  

\begin{deluxetable*}{ccc}
\tablenum{1}
\tablecaption{Focal-plane module components are given below for clarity.  Each UFM has two halves, fed by a pair of coaxial cables. The LF focal-plane module has fewer detectors and will use a simplified version of the readout assembly. \label{tab:ufm}}
\tablewidth{0pt}
\tablehead{\colhead{Component in UMM/UFM} & \colhead{Quantity per coaxial chain} & \colhead{Quantity per module}}
\startdata
Multiplexing chips    &    14     &    28     \\  
Multiplexing channels &    910    &    1820   \\ 
Optical detectors  &    864   &    1728    \\ 
Dark detectors     &    12 (or 24)   &    36  \\ 
Transmission lines &    1      &    2    \\ 
Flux ramp lines    &    1      &    2    \\ 
TES bias lines    &    6     &    12    \\ 
\hline
\hline
Component in multiplexer chip & Quantity & \\
\hline
Channels            &    65     &       \\
Diagnostic channels    & 1     &         \\
\enddata
\end{deluxetable*}

\subsection{Focal-plane module}
An exploded view of the UFM is shown in Fig.~\ref{fig3}. The UFM is comprised of three parts: the UMM, the detector stack, and the optical coupling. 
Each SO detector array will have 1764 bolometers~\citep{hill2018}  for both the mid-frequencies (MF) (90, 150~GHz) and ultra-high frequencies (UHF) (220, 290~GHz)~\citep{walker_2019,stevens_2019}, arranged in a hex-pack at a 5.3~mm pixel pitch. 
The low frequencies (LF) (30, 40~GHz) will have fewer detectors. 
The MF and UHF use arrays that detect photons via horn-coupled ortho-mode transducers~\citep{simon_2018}, while the LF uses lenslet-coupled sinuous antennas. 

The UFM design - except for the detector parameters and horn profile - is identical across MF and UHF frequency bands and receivers. Thus, the UFM design must comply with constraints from both the SATs and LATR. The efficient filling of the focal plane for the SAT, with seven tessellated UFMs, motivates the development of the cold readout to only occupy the same footprint as the detector arrays. The tight space constraints in the LATR optics tubes dictates the maximum height of the UFM. 

In the UFM, the UMM is vertically placed atop the detector stack, which itself is already paired to the primary optical coupling both mechanically and thermally, the latter via gold wirebonds. 
The components are aligned via a central and an outer dowel pin, which are terminated in the optical coupling (i.e. horns). The detector stack and UMM each have a central pin hole and outer slotted hole. The latter is sized to accommodate the difference in thermal contractions between components when the modules are cooled to the 100~mK operating temperature.

The optical coupling has a mechanical flange that interfaces to the receiver, wherein the focal-plane modules are tiled. 
The TES bias connections along the perimeter of the UMM are wirebonded to the detector wafer, completing the electrical circuits within the module. 
The assembly is enclosed by a superconducting aluminum outer shield, which also provides force on the readout and detector stack through integrated tripods, completing the mechanical packaging.
The aforementioned DC and RF connectors that interface to the cryostat are fed-through and attached to this outer shield. 

\section{Design validation} \label{sec:validation}
\begin{figure*}[t]
\centering
\includegraphics[width=0.8\linewidth]{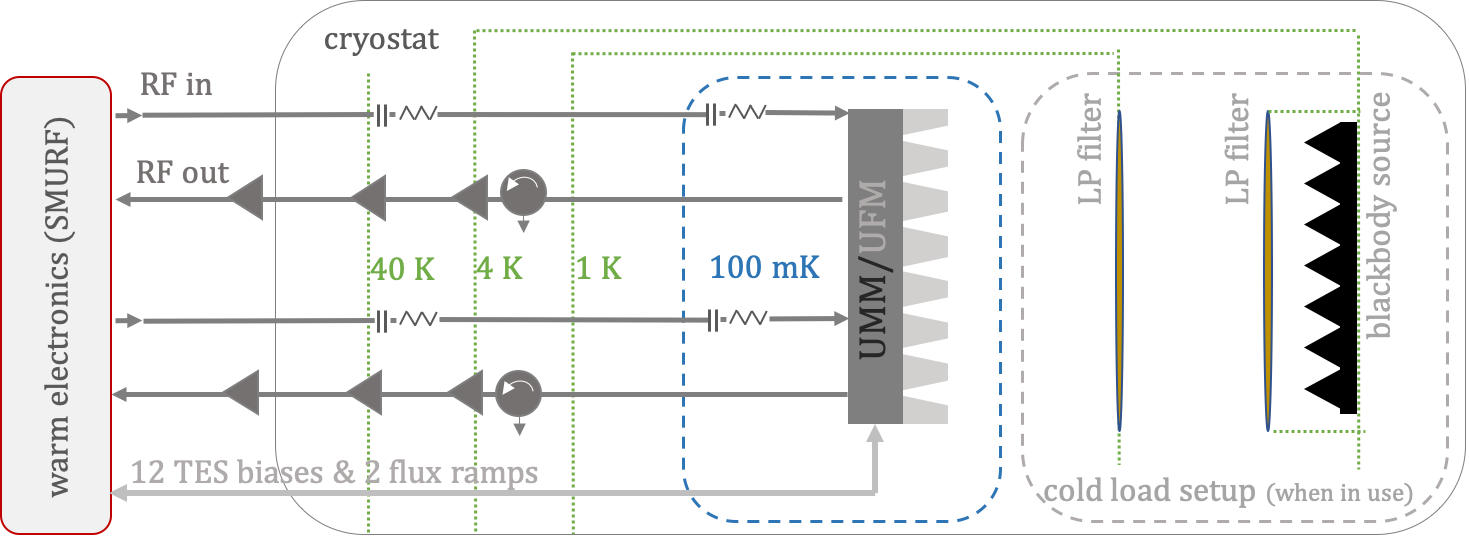}
\caption{Schematic showing the cryogenic readout system and optical testing setup used for design validation. The readout consists of the warm electronics, called SMuRF~\citep{henderson2017}, the RF chain, and the cold multiplexer module. The warm readout generates and reads back the RF probe tones, low-frequency flux ramp signals, and TES bias pairs; the RF chain brings the signal to and from the UMM, the cold mutltiplexer. The two-stage low-noise cold amplification achieves high gain while providing sufficient linearity to drive the 2 $\times$ 910 resonators. Optical testing of the UFM uses a blackbody cold load shown at right. }

\label{fig:readout_overview}
\end{figure*}

To validate the design, we assess the performance of the readout and focal plane modules. The UMM implements the highest TES multiplexing factor to date. Demonstrating its functionality also validates the full system: the complete complement of multiplexer chips, the RF chain and the warm electronics, SMuRF. We first test the multiplexing module in isolation (no TESs), and second when connected to optically coupled TESs. We report measurements of the yield, noise properties, and RF transmission of the multiplexing module in Sec.~\ref{sec:validation_umm}. We then look at the performance when connected to a detector array, reporting the yield, noise properties and detector saturation powers in Sec.~\ref{sec:validation_ufm}. The performance exceeds the science requirements in all cases, as discussed in Sec.~\ref{sec:so_requirements}. 

\subsection{Readout module} \label{sec:validation_umm}
We characterize the multiplexing module in isolation to assess its performance and validate the design choices.  
In this configuration, the UMM is simply mounted to a copper base in lieu of the detector stack and enclosed by the aforementioned aluminum shield. The shield will remain in place for the UFM configuration. 
The UMM is as described in the design section with an exception: one multiplexer chip was replaced by a through CPW. 
The testbed is schematically depicted in Fig.~\ref{fig:readout_overview}. Critically, the two-stage amplification RF chain discussed previously is implemented in the dilution refrigerator. The UMMs are cooled to 100~mK. 

\begin{figure*}
\subfloat[]{%
  \includegraphics[width=0.45\textwidth]{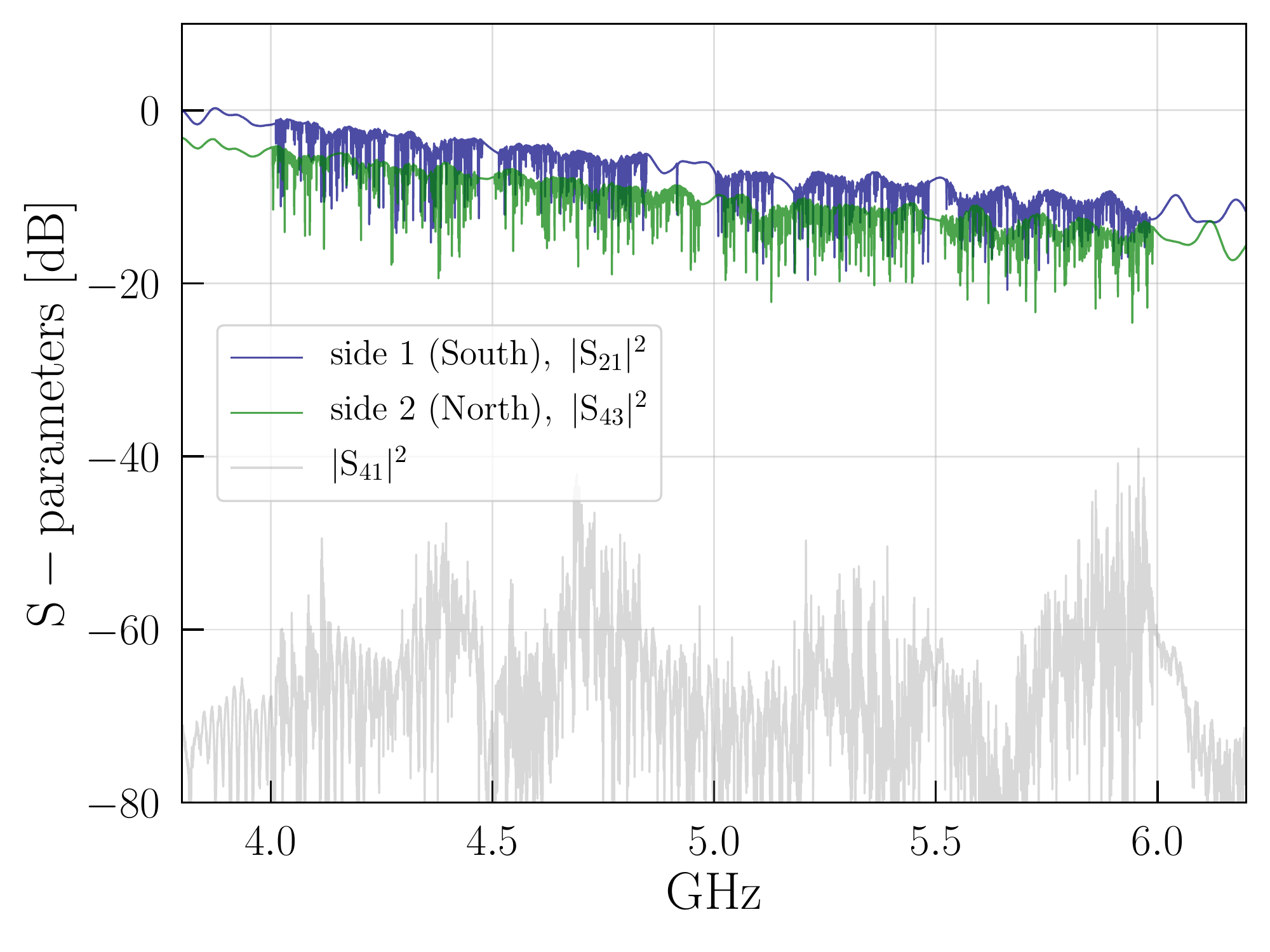}%
}\qquad
\subfloat[]{%
  \includegraphics[width=0.45\textwidth]{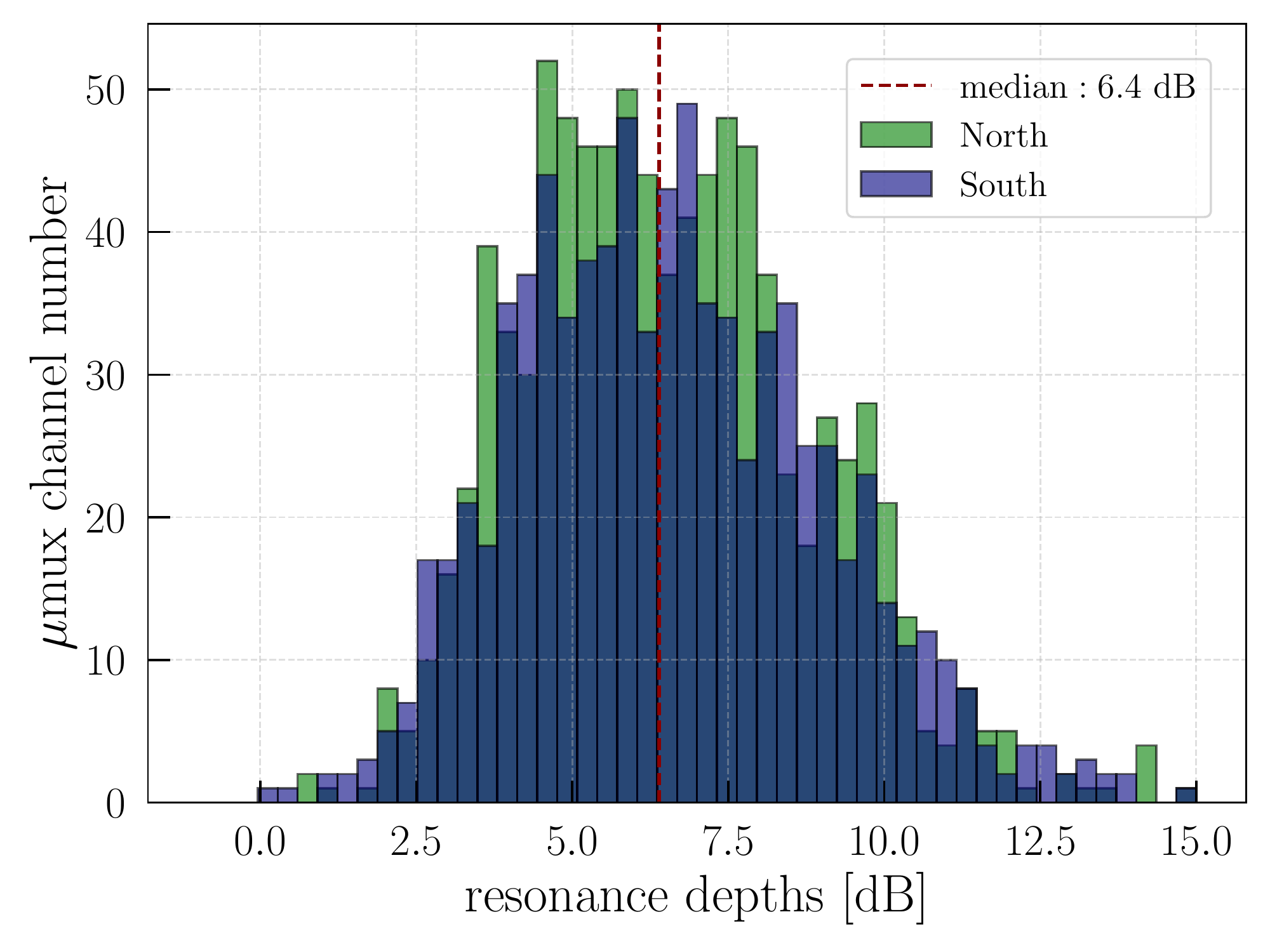}%
}\qquad
\subfloat[]{%
  \includegraphics[width=0.45\textwidth]{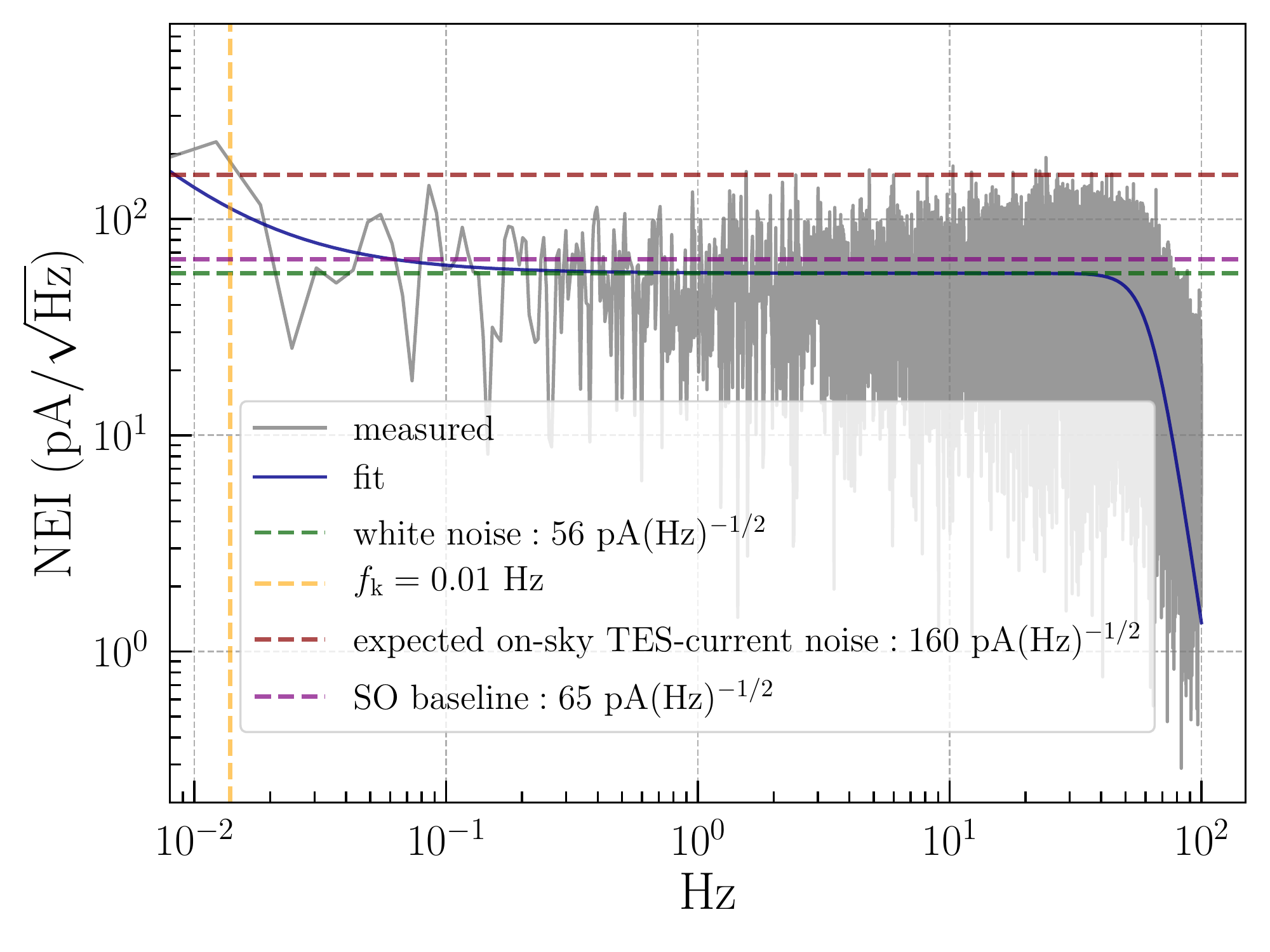}%
}\qquad
\subfloat[]{%
  \includegraphics[width=0.45\textwidth]{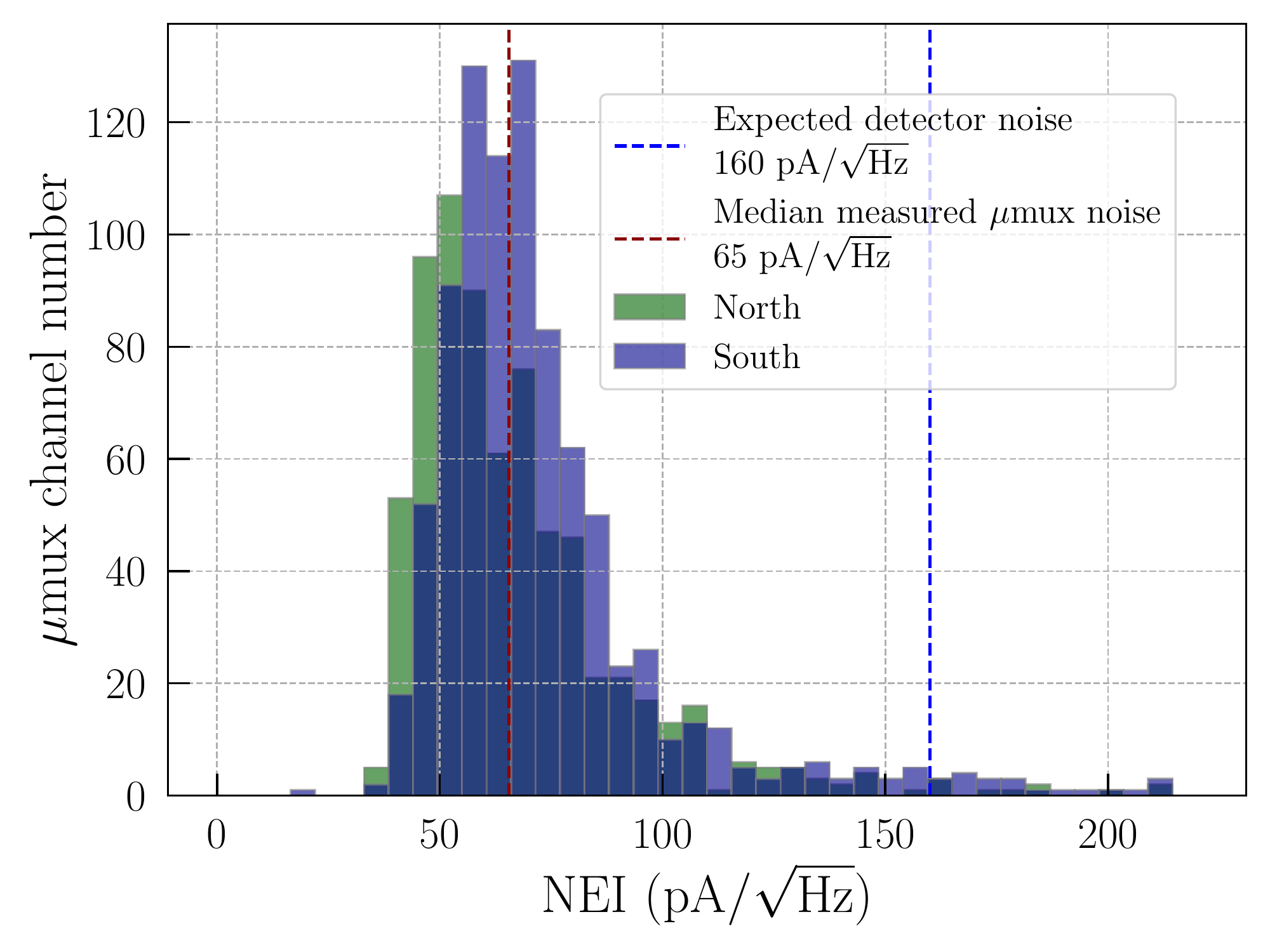}%
}
\caption{{\it Upper left} Transmission through the two sides of a prototype universal $\mu \mathrm{mux}$ module (UMM), each spanning 4--6~GHz. The two halves are referred to as North and South. The isolation between the UMM halves is shown in {\it grey}. The reference plane is at the 300~K shell of the cryostat. The number of readout channels yielded is 1608 for a 92\% yield. 
{\it Upper right} Histogram of the resonance depths. 
{\it Lower left} Noise spectrum of a typical multiplexer channel. Here, the white noise level $S_\mathrm{w}$ ({\it green}) is 56~$\mathrm{pA/\sqrt{Hz}}$ and the upper bound of the frequency knee $f_\mathrm{k}$ is 0.01~Hz. The SO baseline requirement ({\it purple}) is an $f_\mathrm{k}$ of 0.1~Hz with $S_\mathrm{w}< 65~\mathrm{pA/\sqrt{Hz}}$. The expected on-sky TES-current noise ({\it red}) is approximately 160~$\mathrm{pA/\sqrt{Hz}}$ in the 90 and 150~GHz spectral bands.
{\it Lower right} Histogram of the white noise levels of the cold multiplexer, without TESs attached. The array NEI meets the sensitivity necessary to reach the baseline SO forecast~\citep{so_forecast_2019}.}
\label{fig:data}
\end{figure*}

The first performance diagnostic is to quantify the RF transmission. We measure the forward scattering parameter $S_\mathrm{21}$ (and $S_\mathrm{43}$) through the assembly with a standard vector network analyzer.
 Representative scattering parameter are shown in Fig.~\ref{fig:data}. 
We find 1728 resonators, including diagnostic channels, or that 97\% of the resonators yielded. 
We fit a complex model~\citep{kahlil} to the transmission data in order to extract the resonator parameters. This yields the complex resonator coupling quality factor $Q_\mathrm{c}$ and total resonator qualify factor $Q_\mathrm{r}$.
We define $Q_\mathrm{i}$ as  $Q_\mathrm{i}^{-1} = Q_\mathrm{r}^{-1} - \mathrm{Re}(Q_\mathrm{c})^{-1}$; thus $Q_\mathrm{i}$ includes all sources of loss. 
The median $Q_\mathrm{i}$ is {$\sim 7 \times 10^{4}$}. The corresponding median resonance depth is 6.4~dB; this is defined as the difference between the local baseline off-resonance transmission and the resonance minimum.
The median resonator bandwidth $b_\mathrm{r}$ is  {$\sim 80$}~kHz. 
The scattering parameters for inter-line isolation are $S_\mathrm{23}$(and $S_\mathrm{41}$), but we compare these levels to the transmission $S_\mathrm{21}$ (and $S_\mathrm{43}$) level to account for the loses between the calibration referenced plane and the device. We find that this exceeds 20~dB at all but the highest frequencies.  

Second, we measure the noise. 
With the SMuRF warm electronics, both sides of the UMM are multiplexed, meaning all the channels on both halves are read out simultaneously for a multiplexing factor of $2 \times 910$. 
The optimized drive power per resonator at the UMM input is approximately -78~dBm.  
The noise is measured by taking time-ordered-data.
The noise properties are measured by computing the amplitude spectral density for each readout channel from the corresponding time-ordered data and fitting a noise model. The model is parameterized by a low-frequency knee $f_\mathrm{k}^\alpha$, where $\alpha$ is a free index, and the white noise level $w_\mathrm{l}$. The high-frequency roll-off is determined by the anti-aliasing filter. 
The median {input-current-referred} white noise level or noise-equivalent-current (NEI), as fit from this model, of the 1608 $\mu \mathrm{mux}$ channels is $65~\mathrm{pA/\sqrt{Hz}}$ (as shown in Fig.~\ref{fig:data}), with a median  $f_\mathrm{k}$ of $<$ 0.02~Hz.
The noise is much smaller than the expected on-sky TES-current noise, which, for SO, is $160~\mathrm{pA/\sqrt{Hz}}$ in the 90 and 150~GHz spectral bands. 

The final key performance criteria is the yield. 
On the UMM half with a full complement (14) of multiplexer chips and 910 readout channels, the yield is 95\%.
We find a 92\% total yield with 1608 channels across the full module as reported in the plots in Fig.~\ref{fig:data}. 
The 8\% loss can be attributed to the minority of resonators that had collisions in frequency space, fell outside the SMuRF-defined band filters, or were coupled to a SQUID with an anomalous response.  

We configure a separate readout module specifically to measure crosstalk between the two halves. As previously described, the resonator frequencies are repeated twice in each readout module, once on each half. We thus are concerned with measuring the crosstalk between resonators that fall very close to one another in frequency space across the two halves. The crosstalk within an individual multiplexer chip, measured with a `victim and perpetrator' set-up, was shown to be $<0.3\%$~\citep{dober_2020}. Using the same method with the UMM, we choose a 'perpetrator' resonator with the 'victim' resonators on the other half  of the array. The closest resonance frequencies across the two halves fall within one linewidth ($<100$~kHz) of each other. We measure the maximum crosstalk between the channels to be $< 0.1\%$.

\begin{figure*}[htp]
\centering
\begin{tabular}{@{}c@{}}
\subfloat[ \label{fig:vRBd}]{\includegraphics[width=0.4\linewidth]{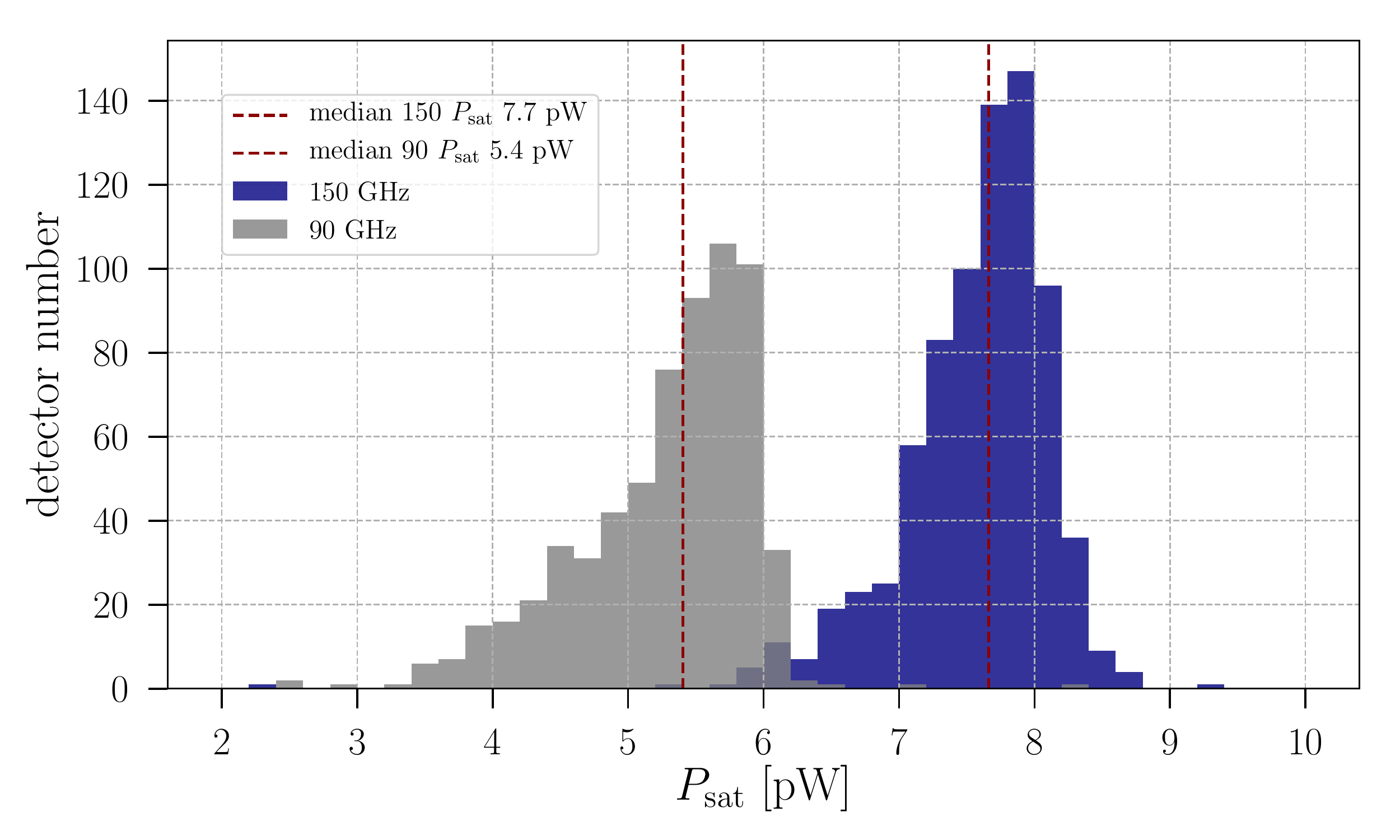}}
\\
\subfloat[ \label{fig:msD}]{\includegraphics[width=0.42\linewidth]{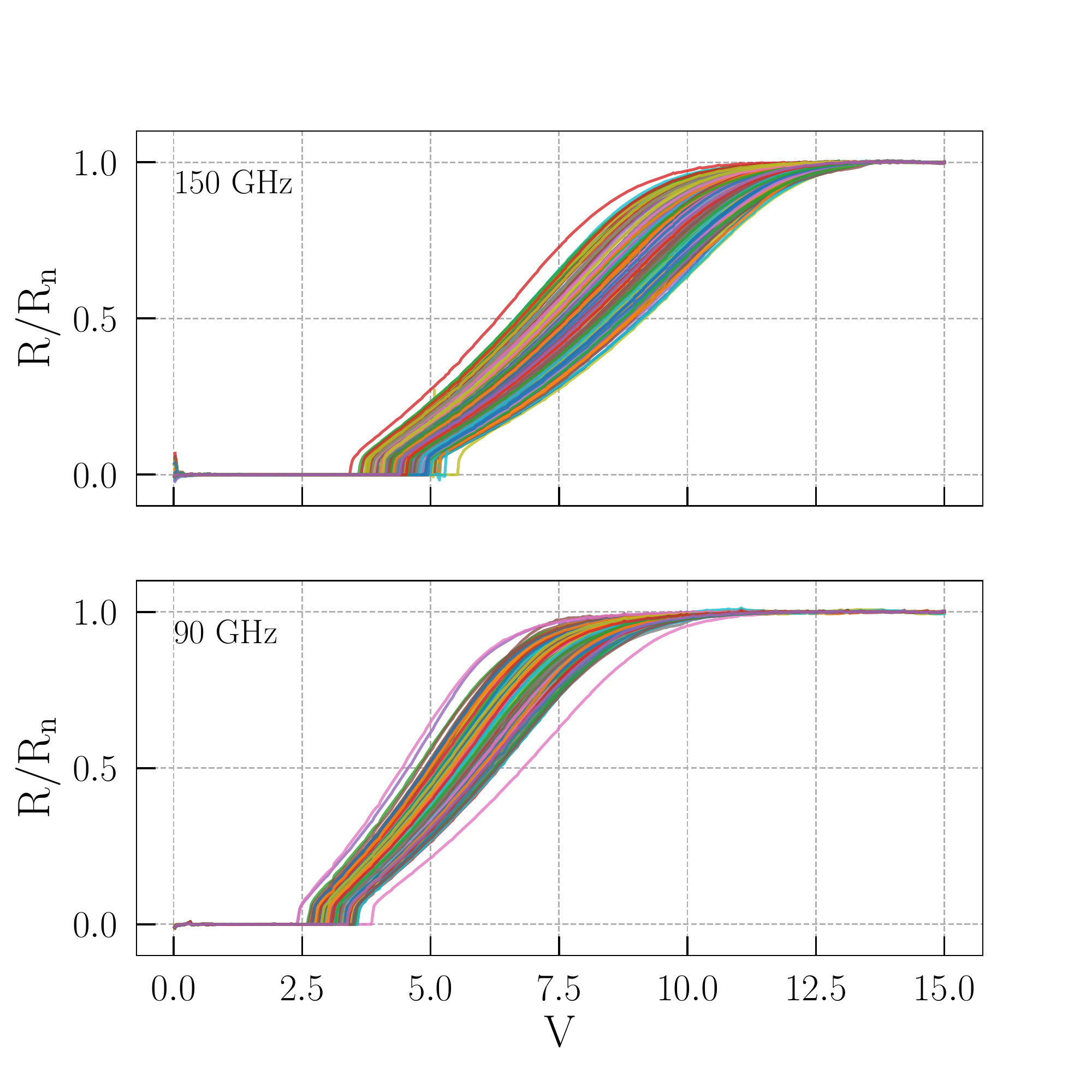}}
\end{tabular} 
\begin{tabular}{@{}c@{}}
\subfloat[ \label{fig:HdRM}]{\includegraphics[width=0.45\linewidth]{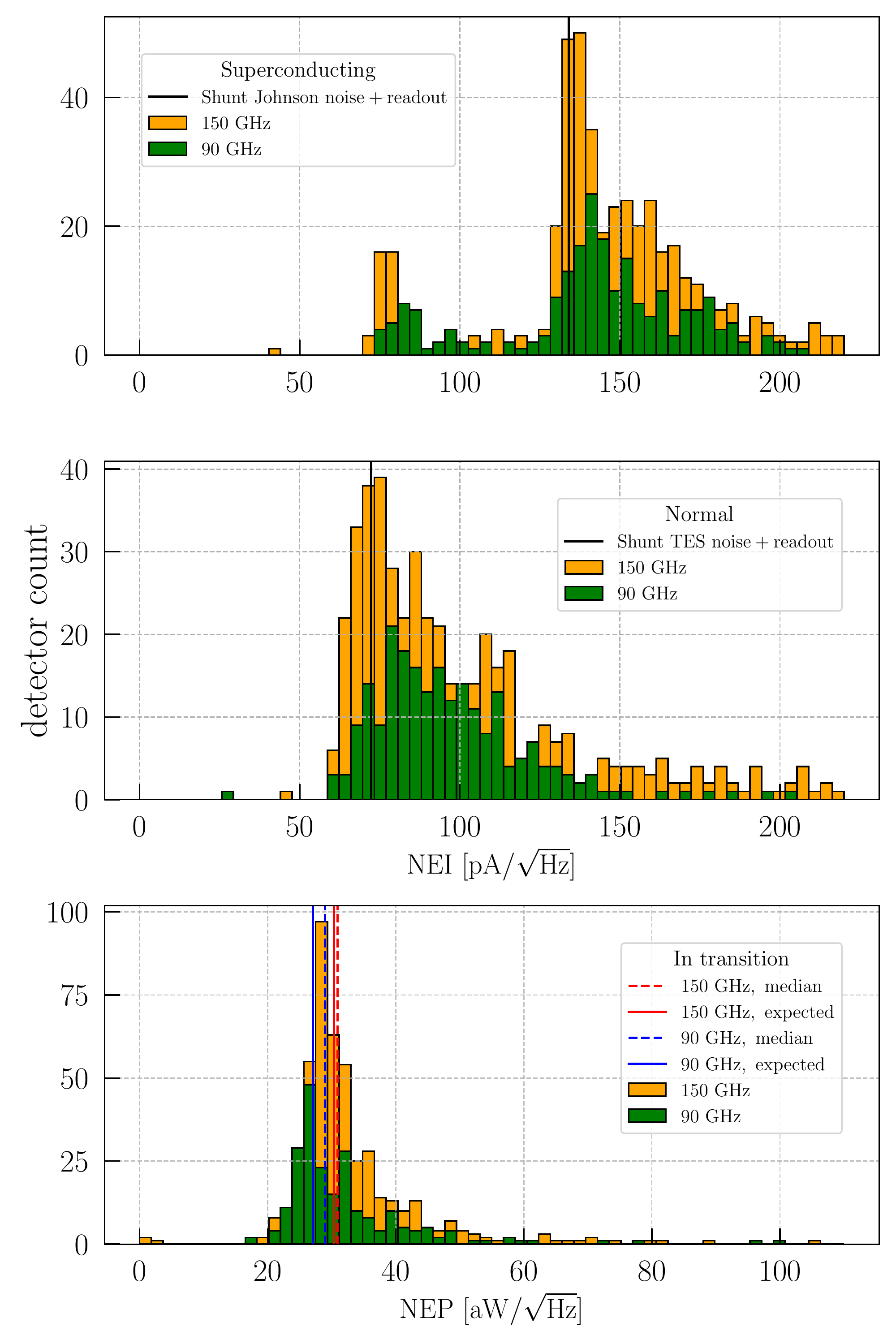}}
\end{tabular}
\caption{{\it Upper left} The saturation powers $P_\mathrm{sat}$ for the detectors sensitive to 90 and 150~GHz are shown in grey and blue, respectively. The median $P_\mathrm{sat}$ for the 90 and 150~GHz detectors are 5.2 and 7.3~pW. The array operability yield is $82\%$. 
{\it Lower left} The normalized resistances of the TESs sensitive to 90 and 150~GHz as they are biased through transition. Each line represents a different detector; the uniformity of the detectors is evident as is the large range over which they can be biased. 
{\it Right} Histograms of the white noise level of the UFM detectors as measured with the $910$ multiplexer; the TESs are in the superconducting, normal, and in-transition states and exposed to an 8.5~K blackbody. The dashed vertical lines show the measured median noise per spectral band and the continuous vertical lines, the expected median noise including the photon noise from the blackbody loading. For both the 150 and 90~GHz detectors, the noise is within 10\% of expectation. Note that (i) the low population in the superconducting histogram is due to the uncoupled readout channels and (ii) the x-axes are in reported in NEI for the top two histograms and in NEP for the bottom. See Appendix~\ref{appendix_noise} for details on the expected noise. }
\label{fig5}
\end{figure*}

\subsection{Focal-plane module} \label{sec:validation_ufm}

We characterize the UFM to demonstrate the readout module's integrated performance and validate the overall design.
The readout module is coupled to a MF prototype SO detector array and horn array. We characterize the UFM in laboratory in two separate configurations: (i) the horn apertures are covered at 100~mK and (ii) the horn apertures are exposed to an 8.5~K blackbody load. In the second configuration, the spectral band is defined by 1~K and 4~K low-pass filters and the on-chip frequency filters fed by the high-pass cylindrical waveguide, which is the input guide. 

First, we re-evaluate aspects of the readout now that it is coupled to a detector array. Significantly, we see no degradation of $Q_{i}$ and thus expect the readout noise to remain the same as when measured independently. The resonator yield also remains unchanged. 

Second, we measure the saturation power $P_\mathrm{sat}$ of the TES bolometers in the dark configuration, with the horn apertures covered at 100~mK.  To do this, the TES biases are stepped over a range in which the detectors are taken from normal to superconducting and the current is measured at each step, mapping out an IV curve. The resulting detector saturation powers, defined at 90\% of the normal resistance, are shown in Fig.~\ref{fig5}. The median saturation powers for the 90 and 150~GHz detectors are 5.2 and 7.3~pW, within 0.2 pW of the SO target for these prototype arrays. The on-specification $P_\mathrm{sat}$ measurements indicate no significant heating on the detector array, which would tend to reduce the observed saturation power. The operable yield in the UFM is $82\%$. 

Third, we measure the noise with the TES bolometers in the normal, superconducting and in-transition states with the horn aperture exposed to the 8.5~K blackbody load. The histograms in Fig.~\ref{fig5} show the measured and expected white noise levels in the three states. In the superconducting, the measured median is 143~$\mathrm{pA/\sqrt{Hz}}$.  The Johnson noise from the shunt resistors, which are measured directly on the routing wafer to be 400~$\mu\Omega$ during screening, is added in quadrature to the previously measured readout noise (as in Sec.~\ref{sec:validation_umm}) to find the total expected noise of 134~$\mathrm{pA/\sqrt{Hz}}$; the noise is within $7\%$ of expectation. With the TES bolometers in the normal state, the measured noise is similarly in agreement with expected noise. The in-transition noise is within 10\% of the expected noise (detailed in Appendix~\ref{appendix_noise}) for both the 90 and 150 GHz detectors.

\section{Simons Observatory requirements and context}
\label{sec:so_requirements}
The science case and forecast for SO is presented in~\citet{so_forecast_2019}. The primary forecast is referred to as the `baseline'. A secondary, more aggressive forecast is referred to as `goal'.  We assess whether the UFM performance meets the baseline metrics, which requires that (i) the readout reduce the observatory mapping speed by less than 18\%, (ii) the readout have a low-frequency knee less than 0.1~Hz and (iii) the module have an end-to-end yield is $>70\%$.  As will be shown in this section, the requirements are met. 

First, we assess the readout measured independently.  The noise due to the readout at full multiplexing will only result in an $8\%$ degradation of total detector noise. This is better than the target level of 10\% needed to achieve the baseline SO sensitivity. The yield exceeds the 78\% readout yield required for SO. Thus, the weighted readout array NEI is better than required.   The measured low-frequency knee, is constrained to below 0.03~Hz, better than the 0.1~Hz requirement. 
These noise measurements also demonstrate that the nonlinearity of the end-to-end readout system is sufficient such that the third-order intermodulation products are small enough not to cause an excess of noise at the full multiplexing factor.

Second, we assess the UFM performance given the desired SO science goals. The in-transition noise for the 90 and 150~GHz detectors are below the increased $30\%$ noise budget allocated in the forecasting. The 82\% end-to-end yield exceeds the baseline requirements. 
In essence, the focal-plane module design is fully functional and exceeds the requirements necessary to meet the science goals. 
 
The readout modules presented in this paper will be used with both the  MF and UHF detector arrays, comprising the MF and UHF UFMs. The LF modules, with $<200$ detectors per array, will use a simplified readout module design of the same footprint. 
The first UFMs will be deployed within a year. 
There will be a total of 49 UFMs across the four receivers. There are four focal planes for the three SATs (two MF, one UHF, and one LF); the three higher frequency focal planes will deploy initially.  The LATR~\citep{xu2020} will host seven optics tubes, each of which has a focal plane with three UFMs, for 21 UFMs in total. The LATR has been designed to hold 13 optics tubes and can host up to 39 UFMs. 

\section{Future outlook} \label{sec:future}
This focal-plane module design has successfully placed 1820 readout channels in a 150~mm diameter assembly with noise subdominant to the expected on-sky detector noise. The performance is better than the SO requirements. 
 This paper shows the feasibility of using $\mu \mathrm{mux}$ in large-scale experiments. Until now, $\mu \mathrm{mux}$ has been in development and deployed on smaller scales as an individual module~\citep{dicker2014} or technology demonstration~\cite{ari_2019}; the results presented here demonstrate the feasibility of employing the technology to unlock large-scale instruments. The focal-plane modules offer an elegant and now proven design not only for SO but also other receivers, many of which are planned for the imminent future~\citep{cmbs4_2020,ccat_2019}. 
 The high-performance design of next-generation, large-scale arrays is a critical area of study not only for millimeter-wave astrophysics but also necessarily for a larger science landscape from x-ray experiments~\citep{yoon2018} to quantum computing~\citep{Rosenberg2019}.

While stressing that the readout system already meets the baseline requirements for SO to achieve the planned science, we can also place the current tests relative to the `goal' performance requirements. We could plan to improve the readout system along three axes. First, we aim to improve the yield, by making incremental changes to the module design and multiplexer chips, such as addressing small shifts in absolute frequency placement. Second, we aim to reduce the per-channel readout noise to be consistently below $<45~\mathrm{pA/\sqrt{Hz}}$. This performance has been demonstrated in preliminary UMMs. The data presented in this paper, which meets baseline requirements, is shown due to the comprehensive end-to-end testing, including in the UFM configuration. This requirement has also been consistently met in the NIST $\mu \mathrm{mux}$ assemblies with multiplexing factors of 512~\citep{henderson2017,dober2018} but in packaging that cannot be coupled to on-sky detectors. The multiplexer chips measured individually and in the aforementioned packaging achieve higher in-situ $Q_\mathrm{i}$. The difference in achieved performance can be attributed to a combination of factors including the packaging-loss due to the overall dimensions. However, we repeat that the $Q_\mathrm{i}$ measured in the current UMM design are sufficient for the SO noise requirements. We also expect there are incremental gains which can be made by further optimizing SMuRF for the UFM. 
Finally, we are looking at increasing the multiplexing factor of the current module to 1820 and read out each module with a single pair of coaxial cables (instead of  two), further decreasing wiring complexity and thermal loading in the receivers. The overall design of the UFM is compatible with an 1820 multiplexer using 1 $\times$ 28 multiplexer chips to cover a 4--8~GHz bandwidth. Both the multiplexer chips and SMuRF have demonstrated their respective capabilities in the 6-8~GHz range.

To conclude, we have presented the design choices and in-lab demonstrations that validate this overall focal-plane architecture.  
We demonstrated the 910 multiplexer with a 95\%  yield in the focal-plane packaging. 
Multiplexing all channels, the median white noise is $< 65~\mathrm{pA/\sqrt{Hz}}$, only a 8\% decrease in sensitivity compared to the expected on-sky TES noise, with the largest number of channels multiplexed to this date. 
With the UMM, we have demonstrated the operation of a SO TES bolometer array within the focal-plane module, the UFM. 
The first UFMs are to be integrated into the SO receivers within the next year. 

\begin{acknowledgments}
This work was supported by a grant from the Simons Foundation (Award $\#457687$, B.K.). SKC acknowledges support from NSF award AST-2001866. YL acknowledges Kavli Institute at Cornell Postdoctoral Fellowship. This document was prepared by the Simons Observatory using the resources of the Fermi National Accelerator Laboratory (Fermilab), a U.S. Department of Energy, Office of Science, HEP User Facility. Fermilab is managed by Fermi Research Alliance, LLC (FRA), acting under Contract No. DE-AC02-07CH11359. ZX is supported by the Gordon and Betty Moore Foundation. 
\end{acknowledgments}

\appendix

\section{Expected noise}\label{appendix_noise}
In this appendix, we present the calculation for the expected noise of the readout and detectors, which is referenced in Sec.~\ref{sec:validation_ufm} and Fig.~\ref{fig5}. The expected noise is calculated for the TES bolometers in three states:  superconducting, normal, and in transition.  

We consider contributions from the readout noise ($\mathrm{ro}$); Johnson noise in the shunt resistance (J,s); Johnson noise in the TES when it is normal (J,n); thermal carrier noise in the TES bolometers (G); and bolometer noise due to photon loading ($\mathrm{\gamma}$). The parenthetical labels are used as subscripts below.  
The $\mu$mux readout architecture converts power fluctuations at each TES bolometer into current fluctuations, which inductively induce changes in the resonator frequency. Ultimately, the signal is read out in radians and referenced to current through the TES, proportional to the mutual inductance between the SQUID and resonator.  The first three noise terms listed above arise as current noise (NEI)  but can be converted to power units (NEP) by noting that in the TES, power fluctuations $\delta P$ are related to current fluctuations $\delta I$ by  the voltage across the TES, $V_\mathrm{TES}$:  $\delta P = \delta I V_\mathrm{TES}.$  Thus, 
$$ \mathrm{NEP} = V_{\rm{TES}} \, \mathrm{NEI}.$$ 
In practice, we use  $V_\mathrm{TES} = (I_\mathrm{bias} - I_\mathrm{TES})R_\mathrm{sh}$, where $I_\mathrm{bias}$ is calculated from the warm, commanded voltage and known wiring resistance leading to the array,  $I_\mathrm{TES}$ is measured as described above, and $R_\mathrm{sh}$ is the shunt resistance, designed and measured to be 400~$\mu\Omega$ in screening.  

The expected  total noise in the three scenarios is:
\begin{enumerate}
\item  Superconducting: $\mathrm{NEI}_\mathrm{s} = ( \mathrm{NEI}_\mathrm{ro}^2 + \mathrm{NEI}_\mathrm{J,s}^2)^{(1/2)}$,
\item  Normal: $\mathrm{NEI}_\mathrm{n} = ( \mathrm{NEI}_\mathrm{ro}^2 + \mathrm{NEI}_\mathrm{J,n}^2)^{(1/2)}$,
\item  In transition: $\mathrm{NEP}_\mathrm{t} = ( \mathrm{NEP}_\mathrm{ro}^2 + \mathrm{NEP}_\mathrm{G}^2 + \mathrm{NEP}_\mathrm{\gamma}^2)^{(1/2)}$.
\end{enumerate}

\subsection{Current noise}
The readout noise is measured  directly in the UMM configuration (no TESes). This level is used as our expected readout noise when calculating the expected total noise in the UFM configuration. The Johnson noise when the TES is superconducting is due entirely to the shunt resistor and is given as  \[\mathrm{NEI_{J,sh}}^2 = \frac{4 k_\mathrm{B} T_\mathrm{b}}{R_\mathrm{sh}},\] where $k_\mathrm{B}$ is the Boltzmann constant and $T_\mathrm{b}$ the array bath temperature, at which the shunt resistors are held. When the TES is normal, with resistance $R_\mathrm{n}$, the Johnson noise is 
given as \[\mathrm{NEI_{J,TES}}^2 = \frac{4 k_\mathrm{B} T_\mathrm{c}}{R_n},\] where $T_\mathrm{c}$ is the TES critical temperature. The contribution from the shunt is negligible as $R_\mathrm{n} >> R_\mathrm{sh}$. The normal resistance is designed to be 8~m$\Omega$;  actual values are extracted from the IV curve measurements.  
 
 \subsection{TES thermal carrier noise}
 With the TES in transition, the bolometer thermal carrier noise is~\citep{mather1982}: 
\[\mathrm{NEP_{G}}^2 = 4 k_\mathrm{B} F T_\mathrm{c}^2 G s_\mathrm{i},\] where $G$ is the bolometer thermal conductance, $s_\mathrm{i}$ is the responsivity, and $F$ is a numerical factor related to the thermal conduction index $n$. We estimate the noise with $F = 1$. The factors of $G$ and $T_\mathrm{c}$ are extracted from fitting data taken from $T_\mathrm{b}$ sweeps: an IV curve is taken as $T_\mathrm{b}$ is stepped from well below to above $T_\mathrm{c}$ and the $P_\mathrm{sat}$ measured at each step. The $P_\mathrm{sat}$ as a function of $T_\mathrm{b}$ is fit to the typical relation~\citep{irwin2005} $\kappa (T_\mathrm{c}^{n} - T_\mathrm{b}^{n})$, where $\kappa$ is a constant, which directly yields $T_\mathrm{c}$ and $n$. $G$ can then be calculated as $G = n \kappa T_\mathrm{c}^{n-1}$. The responsivity is calculated from the local slope from the IV curve~\citep{irwin2005}.   
  
 \subsection{Photon noise}
 
In transition, the photon noise is the dominant noise source. The photon noise is calculated as~\citep{Zmuidzinas2003}
 
 \[\mathrm{NEP_{p}}^2 =  
 2 \int_{\nu_\mathrm{lp}}^{\nu_\mathrm{hp}} 
 \Big[ h \nu p(\nu) B(\nu) +  \big(p(\nu)B(\nu)\big)^{2} \Big] 
\,d\nu. \]
 
 Here, $\nu$ is the photon frequency, which for these paper measurements, is emitted by the cold blackbody ($T = 8 - 20~\mathrm{K}$), and $\nu_\mathrm{lp}$ and $\nu_\mathrm{hp}$ are the frequencies over which the expression is integrated and outside the expected bandpass. The power spectral density $p(\nu)$ can be further expanded  as $p(\nu) = S(T,\nu) \eta \epsilon(\nu)$, in which $S(T,\nu)$ is the standard modified blackbody equation for a single polarization and single mode at temperature $T$, $\epsilon(\nu)$ is the emissivity (in practice, taken to be 1), and $\eta$ is the detector optical efficiency, gathered by measuring $P_\mathrm{sat}$ as a function of stepped $T$ (as in~\cite{choi_2018}) with a prototype detector array of the same design and fabrication run. $B(\nu)$ is the cumulative bandpass, which has multiplicative contributions from the band defining elements: the metal-mesh filters (the spectra are measured separately), low-pass waveguide (simulated), on-chip frequencies filters (simulated) and beam-fill fraction (calculated).

\bibliography{bib}{}
\bibliographystyle{aasjournal}

\end{document}